# A Novel Method for Intrusion Detection System to Enhance Security in Ad hoc Network

[1] Himani Bathla, [2] Kanika Lakhani

**Abstract**— The notion of an ad hoc network is a new paradigm that allows mobile hosts (nodes) to communicate without relying on a predefined infrastructure to keep the network connected. Most nodes are assumed to be mobile and communication is assumed to be wireless. The mobility of nodes in an ad-hoc network means that both the population and the topology of the network are highly dynamic. It is very difficult to design a once-for-all intrusion detection system. A secure protocol should atleast include mechanisms against known attack types. In addition, it should provide a scheme to easily add new security features in the future. The paper includes the detailed description of Proposed Intrusion Detection System based on Local Reputation Scheme. The proposed System also includes concept of Redemption and Fading these are mechanism that allow nodes previously considered malicious to become a part of the network again. The simulation of the proposed system is to be done using NS-2 simulator.

——————————— ◆ ———————————

## 1 INTRODUCTION

A mobile ad hoc network (MANET) is a self-configuring network that is formed automatically by a collection of mobile nodes without the help of a fixed infrastructure or centralized management. Each node is equipped with a wireless transmitter and receiver, which allow it to communicate with other nodes in its radio communication range. In order for a node to forward a packet to a node that is out of its radio range, the cooperation of other nodes in the network is needed; this is known as multi-hop communication. Therefore, each node must act as both a host and a router at the same time. The network topology frequently changes due to the mobility of mobile nodes as they move within, move into, or move out of the network. There are both passive and active attacks in MANETs. For passive attacks, packets containing secret information might be eavesdropped, which violates confidentiality. Active attacks, including injecting packets to invalid destinations into the network, deleting packets, modifying the contents of packets, and impersonating other nodes violate availability, integrity, authentication, and non-repudiation. Proactive approaches such as cryptography and authentication were first brought into consideration, and many techniques have been proposed and implemented. However, these applications are not sufficient. If we have the ability to detect the attack once it comes into the network, we can stop it from doing any damage to the system or any data. Here is where the intrusion detection system comes in.

  Intrusion detection can be defined as a process of monitoring activities in a system, which can be a computer or network system. The mechanism by which this is achieved is called an intrusion detection system (IDS). An IDS collects activity information and then analyzes it to determine whether there are any activities that violate the security rules. Once an IDS determines that an unusual activity or an activity that is known to be an attack occurs, it then generates an alarm to alert the security administrator. In addition, IDS can also initiate a proper response to the malicious activity.

  Many historical events have shown that in-

————————————————
- *Himani Bathla is with Gurgaon Institute of Technology & management, Gurgaon.*
- *Kanika Lakhani is with Institute of Technology & Management, Gurgaon.*

trusion prevention techniques alone, such as encryption and authentication, which are usually a first line of defense, are not sufficient. As the system become more complex, there are also more weaknesses, which lead to more security problems. Intrusion detection can be used as a second wall of defense to protect the network from such problems. If the intrusion is detected, a response can be initiated to prevent or minimize damage to the system.

Some assumptions are made in order for intrusion detection systems to work. The first assumption are that user and program activities are observable. The second assumption, which is more important, is that normal and intrusive activities must have distinct behaviors, as intrusion detection must capture and analyze system activity to determine if the system is under attack.

Intrusion detection [1] can be classified based on audit data as either host based or network-based. A network-based IDS captures and analyzes packets from network traffic while a host-based IDS uses operating system or application logs in its analysis. Based on detection techniques, IDS can also be classified into three categories as follows.

**Anomaly detection systems**: The normal profiles (or normal behaviors) of users are kept in the system. The system compares the captured data with these profiles, and then treats any activity that deviates from the baseline as a possible intrusion by informing system administrators or initializing a proper response.

**Misuse detection systems**: The system keeps patterns (or signatures) of known attacks and uses them to compare with the captured data. Any matched pattern is treated as an intrusion. Like a virus detection system, it cannot detect new kinds of attacks.

**Specification-based detection**: The system defines a set of constraints that describe the correct operation of a program or protocol. Then, it monitors the execution of the program with respect to the defined constraints.

## 2 INTRUSION DETECTION IN MANET

Many intrusion detection systems have been proposed in traditional wired networks, where all traffic must go through switches, routers, or gateways. Hence, IDS can be added to and implemented in these devices easily. On the other hand, MANETs do not have such devices. Moreover, the medium is wide open, so both legitimate and malicious users can access it. Furthermore, there is no clear separation between normal and unusual activities in a mobile environment. Since nodes can move arbitrarily, false routing information could be from a compromised node or a node that has outdated information. Thus, the current IDS techniques on wired networks cannot be applied directly to MANETs. Many intrusion detection systems have been proposed to suit the characteristics of MANETs, some of which will be discussed in the next sections

## 3. INTRUSION DETECTION TECHNIQUES FOR NODE COOPERATION IN MANET

There are several proposed techniques and protocols to detect such misbehavior in order to avoid those nodes, and some schemes also propose punishment as well:

Watchdog and Path rater: A watchdog identifies the misbehaving nodes by eavesdropping on the transmission of the next hop. A path rater then helps to find the routes that do not contain those nodes. In DSR, the routing information is defined at the source node. This routing information is passed together with the message through intermediate nodes until it reaches the destination. Therefore, each intermediate node in the path should know who the next hop node is.

Reputation Based Schemes: Reputation systems are used in many areas of electronic transactions, such as eBay and Amazon. Reputation mechanisms are applied to wireless mobile ad hoc network to address threats arising from uncooperative nodes. They rely on neighbor monitoring to dynamically assess the trustworthiness of neighbor nodes and excluding untrustworthy nodes. Several reputation systems have been proposed to mitigate selfishness and stimulate cooperation in mobile ad hoc network, including: CONFIDANT, CORE, OCEAN

## 4 Proposed Intrusion Detection System

It is a simple reputation based scheme, called Local Reputation System to mitigate misbehavior and enforce cooperation. Different from global reputation Schemes such as CONFIDANT and CORE, our solution uses local reputation only. Each node only keeps the reputation values of all its one hop neighbors. Local Reputation System addresses the problem of node cooperation in self organized ad hoc networks. In these networks, nodes may not belong to single authority and don't have common goals. By Self Organizing mean that regular function of network depends on End Users operation. In this Trust is associated with its reputation value. There are three trust levels and we use a trust value, T, to represent the trustworthiness of a node. A node A considers another node B either

- trustworthy, with T = 1,
- untrustworthy, with T = -1, or
- trustworthy undecided, with T = 0

A trustworthy node is a regular (well-behaved) node that can be trusted. An untrustworthy node is a misbehaved node and should be avoid. A node with undecided trustworthiness is usually a new node in the neighborhood. It may be a regular or a misbehaved node, depending on its future performance.

Every node keeps a reputation table, which associates a reputation value with each of its neighbors. It updates the reputation table based on direct observation only. No global reputation value is calculated, and no indirect reputation message is distributed. Reputation values R are between a range Rmin < R < Rmax, and there are two threshold, Ru > Rmin for untrustworthy and Rt < Rmax for trustworthy. For a node N with reputation value R and trust value T,

- T = 1 (N is trustworthy), if Rt < R <Rmax,
- T = -1 (N is not trustworthy), if Rmin <R < Ru,
- T = 0 (N is trustworthy undecided), if Ru <R < Rt.

A new node, either a node that just entered the network or a node that has moved to a new neighborhood, will be assigned a reputation value between Ru and Rt because its trustworthiness is unknown.

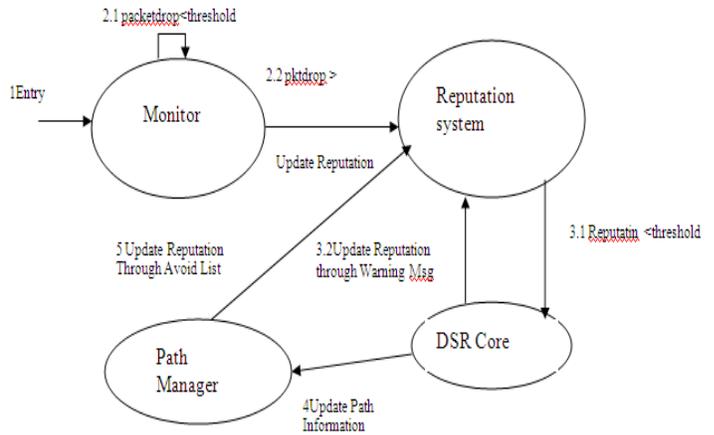

Fig1. IDS based on Reputation

**Monitor**

The Monitor, holds the responsibilty of monitoring activities in the Neighborhood using PACKs (Passive ACKnowledgements) which have been provided as a feature in the DSR protocol specifications as Promiscuous Mode. Every node registers all the data packets sent by it to next node and when it receives packets in promiscuous mode, it matches those to the queue of registered packets present in its buffer. After a fixed time interval -termed as the Timing Window, nodes make a log of number of packets for which they haven't received acknowledgment in the form of PACK and communicate it to the reputation manager. In existing Reputation systems every packet is kept waiting for its PACK for a fixed time interval, in contrast we have used the concept of Timing Window, which gives us flexibility of checking timeout on fixed intervals (that is after every second) rather checking it on basis of each individual packet's timeout. Monitor maintains a log of activity of next neighbor for each Window and sends it to Reputation manager.

**Reputation Manager**

Reputation System assigns and maintain reputation of different nodes.Each node maintains only reputation value of its one hop neighbor. reputa-tion of any node can change by three means:
a. By Self observation
b. WARNING Message, issued by neighboring nodes
c. Avoid List, appended to the RREQ/RREP header.

All three ways have associated reputation weights with them, giving maximum weight to self observation. The reputation is updated after every time window. A node may be tagged as Normal, Suspicious or Malicious depending on reputation associated with it .After each time window Reputation system receives activity log of next hop neighbor from monitor with number of packets for which it does not receive PACK, called as Missing or Dropped Packets. The number of missing is then compared with Malicious Drop Threshold and if it is lesser, then reputation manager gives positive performance else negative. A node is declared malicious only through self observation. When ever any node has a reputation in Malicious category ant that also receive s any new warning message or avoid list ,system performs Trace Test , a test designed to check authentic of node .In this test deciding node generates a fake data packet and forward it to node in question. If next node forward it successfully ,then system gives it a performance appraisal and clears its account else node is declared as Malicious. If the node is actually malicious one then it drop the test packet and Monitor shall report its activity to reputation manager.

**Calculation And Updation of Reputation Value**:
Local Reputation System uses local reputation value, where each node maintains only reputation values of its one hop neighbors. The reputation value is updated based only on its direct observation of the neighbors, no second hand reputation information exchanged and integrated .Suppose a Node A behaves regularly. Everytime it forward a message ,its one hop neighbor observe its Normal behavior and increase its reputation value by say w. $R_x(A)$ mean Reputation Value of node A in node x reputation table.

$$R_x(A) = R_x(A) + w$$

Suppose M is a malicious node and drops the message from its previous node, N. Then there are several different cases. We use the figures below to illustrate each of them. In these figures, a circle describes the one-hop neighborhood of a node, a solid vector describes the forwarding of a message, a dotted vector describes the forwarding of a trace, a dashed vector describes non forwarding (either a message or a trace).

Case 1: N Sent a message to M but M failed to forward it

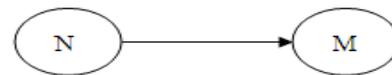

Fig 2. N Sent a message to M but M failed to forward it

For every node X in, which is a one-hop neighbor of both N and M, X detects the misbehavior and reduces the reputation value of M by y, where y > z. That is,

$$R_x(M) = R_x(M) - y$$

Case 2 M has forwarded the message but not forwarded trace.

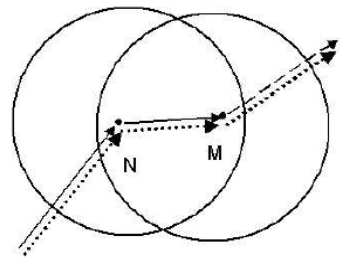

Fig 3.M has forwarded the message but not forwarded trace.

For every node X, which is a one-hop neighbor of M, but not a one-hop neighbor of N, X realizes that M has not forwarded the message when it gets the trace, then X will reduce M's reputation value by y. Thus, all one-hop neighbors of M reduce its reputation by y. That is,

$$R_x(M) = R_x(M) - y, X \in N(M)$$

Case 3: M has not forwarded the message, and has also dropped the trace.

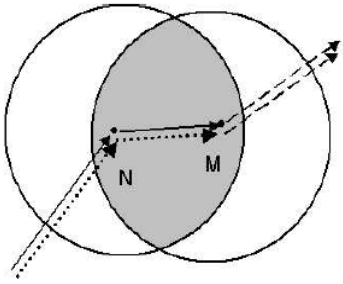

Fig 4. M has not forwarded the message, and has also dropped trace.

$$Rx(M) = Rx(M) - t$$

**Path Manager**
The path manager performs trivial path management functions in collaboration with DSR core. Path ranking is done according to path priority formula. Updating path-cache on various events such as those when new nodes are declared a malicious or a malicious node is taken back in network; taking decision on receiving route request or traffic from malicious nodes are a few responsibilities of the path manager. Concept of avoid list has been added to path manager, which is a list of malicious nodes that a certain node possesses and is appended to the RREQ header whenever a route request is issued by some node. Nodes which find themselves in avoid list do not process the packet and may simply drop it. During RREP, only a path with clean nodes is preferred over those containing suspicious nodes and malicious nodes. Replies from such nodes are also dropped and nodes do not process request and/or forward data packets from such nodes. If during traffic flow, a new node is declared malicious, then, all paths containing it are deleted from route cache and a route error is generated, stating that their link to the destination node has been broken.

**Redemption and Fading**
Redemption and Fading are introduced in our design to allow nodes previously considered malicious to become apart of the network again. MANETs run on cooperation and collaboration of peer nodes and no one gets benefited without cooperating with each other. Trace test is crucial for nodes in suspicious category and node may fail this test due to various reasons like transient link failures, congestion or resetting of the network interface etc. and once they fail this test, they are declared as malicious. To account for these problems, our system uses the fading mechanism. After a certain inactivity period the reputation of a node is improved by a certain predefined fading rate and finally the node is moved from the malicious list to middle of suspicious category. But, the node is not given neutral rating so that if the node again misbehaves then it is immediately put in malicious list and all transactions through that node are blocked. Here, inactivity period means no appearance in any WARNING messages or avoid list.

Following algorithms give a concise idea of the Route Discovery phase, Monitoring mode and Trace test feature of our system. The algorithm for Route Discovery phase is:

```
SENDER
1 Generate RREQ Packet
2 Pack Malicious List in RREQ Header as Avoid List
3 Propagate Request
 OTHER NODES
4 if (Own name present in Avoid List) then
5 Drop Request
6 else
7 Scan Avoid List
8 Update Node's Reputation
9 Append its own malicious list to RREQ header avoiding repetition
10 if (Node is same as Destination in RREQ) then
11 Prepare Reply
12 else
13 Add itself in route and propagate
14 end if
15 end if
```

The above algorithm presents a node's behavior during route establishment phase. Sender of the RREQ just initiates the route discovery process with avoid list of malicious nodes packed in the RREQ packet header. The remaining nodes after receiving such requests process the avoid list attached in the received RREQ header. If a matching entry is found for their own name in the list, the node drops the request. Otherwise, the reputation of the other nodes present in the avoid list is updated. If the receiving node is the destination for which the RREQ has been sent, then it prepares a route reply else it appends its own malicious list in the header to the existing avoid list avoiding repetitions and propagates the route request. Below is the algorithm For Monitor Mode Phase:

```
MONITOR MODE
Self Observation-
1 if (Performance is below normal Threshold) then
2 Negative reputation update
3 else
4 Positive reputation update
5  if (reputation is above 0) then
6 SET reputation = 0
7 end if
8 end if
WARNING MESSAGE PROPAGATION
9 if (WARNING MSG && NEIGHBOR) then
10 if (Reputation below Suspicious Threshold) then
11 Perform Trace Test
12 if (Trace Test is Passed) then
13 Assign normal reputation
14 else
15 declare as Malicious
16 spread Warning Message
17 end if
18 end if
19 else
20 decrease reputation
21: end if
```

Now, we'll discuss the algorithm for trace test

```
Trace Test
1 Identify target Node
2 Generate fake data packet with route via target node
3 Send packet to target node and wait for its PACK
4 if (PACK is found) then
5 test Passed
6 Set reputation to default
7 else
8 test Failed
9 Declare node as malicious and broadcast Warning message
10 end if
```

Trace test is designed specifically for immediate neighbors to test whether a particular node is malicious or not and is only performed on nodes in suspicious state. In this test a dummy data packet with time to live (TTL) equal to 2 is sent to a node in question via last known route through that node. The sender node overhears traffic of the node in question in promiscuous mode. If the node on which trace test is being performed successfully forwards the test packet to next hop then its reputation is set to default. In case it fails, then it is immediately put into malicious category and a warning message is broadcasted in the immediate neighborhood

In case, the dummy packet is genuinely dropped because of bad channel conditions the node may be classified as malicious. However, it still has an opportunity to become a part of network again through redemption and fading mechanism, as explained earlier. This is done because the system only trusts first hand information for putting a node into malicious category, thus, giving self observation the highest weightage. The weight age assigned to warning message and avoid list is comparatively less than self observation.

## 5 CONCLUSION

Mobile adhoc networks have a number of significant security issues which cannot be solved alone by simple IDS. In this, we have to critically examine the existing systems and outline their strength and shortcomings. We have opted an approach for our system in terms of mode of information propagation among nodes. The goal is to design a system incorporating the best traits of all existing systems without incurring extra routing overhead. Trace test and Timing window and Redemption and Fading mechanism are some new concepts that have to be introduced in this system. We will discuss various attacks in ad hoc network. In Future the simulation of the proposed work will be done using NS-2.

**First A. Author**
Himani Bathla is M.Tech in Computer Science from Careers Institute of Technology & management, Faridabad and is currently working as Lecturer in Gurgaon Institute of Technology & management, Gurgaon.

**Second B. Author**
Kanika lakhani is pursuing M.Tech in Computer Science from Institute of Technology & Management, Gurgaon.